\documentclass[iop]{emulateapj}
\usepackage{amsmath, amssymb}
\usepackage{amsfonts}

\newcommand   *{\Msun}       {{\mathrm{M}_{\odot}}}
\renewcommand *{\d}          {\mathrm{d}}
\newcommand   *{\peri}[1]    {#1_{\scriptscriptstyle -}}
\newcommand   *{\apo}[1]     {#1_{\scriptscriptstyle +}}

\begin{document}

%%%%%%%%%%%%%%%%%%%%%%%%%%%%%%%%%%%%%%%%%%%%%%%%%%%%%%%%%%%%%%%%%%%%%%%%%%%%%%%%
\title{Black Hole Foraging: Feedback Drives Feeding}
\author{Walter Dehnen\altaffilmark{1} and Andrew~King\altaffilmark{1}}
\altaffiltext{1} {Theoretical Astrophysics Group, University of
	Leicester, Leicester LE1 7RH, U.K.; wd11@leicester.ac.uk, ark@astro.le.ac.uk}

\begin{abstract}
	We suggest a new picture of supermassive black hole (SMBH) growth in galaxy centers. 	Momentum--driven feedback from an accreting hole gives significant orbital energy 		but little angular momentum to the surrounding gas. Once central accretion drops, 		the feedback weakens and swept--up gas falls back towards the SMBH on
	near--parabolic orbits. These intersect near the black hole with partially opposed
	specific angular momenta, causing further infall and ultimately the formation of a 	small--scale accretion disk. The feeding rates into the disk typically exceed 			Eddington by factors of a few, growing the hole on the Salpeter timescale and 			stimulating further feedback. Natural consequences of this picture include (i) the
	formation and maintenance of a roughly toroidal distribution of obscuring matter 		near the hole; (ii) random orientations of successive accretion disk episodes; (iii) 	the possibility of rapid SMBH growth; (iv) tidal disruption of stars and close 		binaries formed from infalling gas, resulting in visible flares and ejection of 		hypervelocity stars; (v) super--solar abundances of the matter accreting on to 		the SMBH; and (vi) a lower central dark--matter density, and hence annihilation 		signal, than adiabatic SMBH growth implies. We also suggest a simple sub--grid 		recipe for implementing this process in numerical simulations.
\end{abstract}


%%%%%%%%%%%%%%%%%%%%%%%%%%%%%%%%%%%%%%%%%%%%%%%%%%%%%%%%%%%%%%%%%%%%%%%%%%%%%%%%
\section{Introduction}
The relation between supermassive black holes (SMBHs) and their host galaxies
is a major theme of current astrophysics. The scaling relations
\citep{FerrareseMerritt2000, GebhardtEtAl2000, HaeringRix2004} between the
SMBH mass $M$ and the velocity dispersion $\sigma$ and mass
$M_{\mathrm{bulge}}$ of the host spheroid strongly suggest that the hole's
enormous binding energy affects the host in important ways. A credible picture
of this process is gradually emerging \citep[e.g.][]{SilkRees1998,Fabian1999,
  King2003,King2005,ZubovasKing2012:ApJ}. But we are still far from a
deterministic theory of SMBH--galaxy coevolution, because we have no cogent
picture of how the host affects the hole, i.e.\ of what causes SMBH mass
growth. We know that this must largely occur through accretion of gas: the
\cite{Soltan1982} relation implies that mass growth produces electromagnetic
radiation with accretion efficiency $\eta \simeq 0.1\times$ rest--mass energy,
at least at low redshifts. This rules out dark--matter accretion as a major
contributor, and direct accretion of stars through tidal disruption is
inefficient \citep{FrankRees1976}.

Because all gas has angular momentum, accretion on to the hole at the smallest
scales must be through an accretion disk. But these scales must indeed be
small: the viscous timescale
\begin{equation}
\label{eq:t:visc}
  t_{\mathrm{visc}}=\frac{1}{\alpha}\left(\frac{R}{H}\right)^2
  \left(\frac{R^3}{GM}\right)^{1/2}
\end{equation}
approaches a Hubble time at scales of only a few times $0.1\,$pc if the
accreting gas can cool, so that the disk aspect ratio $H/R\ll1$
(e.g.\ \citealt{KingPringle2006, KingPringle2007}; $\alpha\lesssim1$ is the
standard \cite{ShakuraSunyaev1973} viscosity parameter). However, if
$M_{\mathrm{disk}}/M\gtrsim(H/R)\sim0.003$, the disk is self-gravitating and
forms stars instead of accreting. Therefore, for efficient black-hole growth,
$\gtrsim10^{2-3}$ individual accretion events are required, each of which
contributes only a small fraction of $M$ and lasts $\lesssim10^6$\,yr \citep{KingPringleHofmann2008}, implying that the accretion disks have radii $R_{\mathrm{disk}} \lesssim 0.003$\,pc. Yet the gas that the hole must eventually accrete, which can be of order $10^{8-9}\Msun$, must occupy a far larger region $R_{\mathrm{gas}} \sim 10-100\,$pc.

So the missing element in current treatments is a connection between these
scales, telling us how gas falls from a region of size $R_{\mathrm{gas}}$ to
make a succession of disks at scales $\sim R_{\mathrm{disk}}$. In numerical
simulations of galaxy evolution, \cite{Bondi1952} accretion is a popular
choice, but has several critical drawbacks.

Two of these are crucial. The first is that in reality all gas has significant
angular momentum, and so cannot fall in radially, in the way envisaged for
Bondi accretion. Angular momentum is the main barrier to accretion. However
since $R_{\mathrm{gas}}\lesssim$ scale height of the ISM, the cold gas in this
region is probably not in large--scale rotation,
i.e.\ has a distribution of (partly) opposing angular momenta with a
small net angular momentum. Therefore, a way of cancelling these opposing angular momenta would greatly enhance accretion.

A second serious problem in using the Bondi formula is its implication that
gas falls towards the black hole because of the destabilizing influence of its
gravity. But the hole's mass is so small compared to that of even a small
region of the galaxy that this is implausible. As we remarked above, the
property of the hole which is highly significant for the galaxy is not its
mass $M$, but its binding energy $\eta c^2M$, where $\eta\simeq 0.1$. In mass
terms, the hole is typically only one part in about $10^{-3}$ of the galaxy
bulge stellar mass $M_{\mathrm{bulge}}$ \citep{HaeringRix2004}. But for
binding energies the situation is reversed: a hole of mass $10^8\Msun$ has
$\eta c^2M \sim 10^{61}$~erg, while the bulge binding energy is
$\sim\sigma^2M_{\mathrm{bulge}} \sim 10^{58}$~erg for a typical velocity
dispersion $\sigma \simeq 200~{\rm km\,s^{-1}}$ (this disparity is even bigger
for smaller SMBH if these follow the scaling relations).

This suggests that the cause of black hole accretion ultimately involves
its effects on the galaxy, i.e. feedback. We already know quite a lot about
black--hole feedback in galaxies, and how it produces the SMBH--galaxy scaling relations. What is important for our purposes here is that the feedback is carried by quasi--spherical winds driven by radiation pressure; these are detected via blue--shifted X--ray iron absorption lines \citep[e.g.][]{PoundsKingPageOBrien2003,
  PoundsReevesKingEtAl2003, TombesiSambrunaEtAl2010,
  TombesiSambrunaEtAl2011}. The winds have momentum scalars $\dot
M_{\mathrm{out}}v \simeq L_{\mathrm{Edd}}/c$, where $L_{\mathrm{Edd}}$ is the
Eddington luminosity of the hole, $\dot{M}_{\mathrm{out}}$ is the wind outflow
rate, and $v\sim\eta c$ its velocity \citep{KingPounds2003}. The winds
interact with the host galaxy by shocking against its interstellar gas, giving
initial postshock temperatures $\sim10^{10}$\,K. While the SMBH is growing,
these shocks lie close to the hole. Here the much cooler ($\sim10^7$\,K)
radiation field produced by accretion removes most of the shock energy through
the inverse Compton effect \citep{King2003}.

So only the wind ram pressure, i.e. the momentum rate $L_{\mathrm{Edd}}/c$
mentioned above, is communicated to the host ISM (these are called
`momentum--driven' flows). This thrust can push the host ISM only modestly
outwards, and is apparently unable to prevent the hole from growing. But once
the hole mass reaches the $M{-}\sigma$ scaling relation, i.e.
\begin{equation} \label{eq:M:sigma}
  M=M_{\sigma}=\frac{f_{\mathrm{g}}\kappa}{\pi G^2}\sigma^4
\end{equation}
with $f_{\mathrm{g}}$ the local gas fraction,
the wind shocks are able to move far away from the hole \citep{King2003,King2005}, beyond the critical radius $R_{\mathrm{cool}}\sim0.5\,$kpc where the radiation field of the accreting black hole becomes too dilute to cool the shocked wind. This now expands
adiabatically (`energy--driven' flow), sweeping the host ISM before it at high
speed ($\sim1000\,$km\,s$^{-1}$) and largely clearing the galaxy bulge of gas
\citep{ZubovasKing2012:ApJ}. This terminates black hole growth, leaving the
hole near the mass~(\ref{eq:M:sigma}).

%%%%%%%%%%%%%%%%%%%%%%%%%%%%%%%%%%%%%%%%%%%%%%%%%%%%%%%%%%%%%%%%%%%%%%%%%%%%%%%%
\section{Feedback causes feeding}
This sequence shows that the growth of the supermassive black hole towards the
$M{-}\sigma$ relation is characterized by quasispherical momentum--driven
outflow episodes which push the interstellar gas out, but do not unbind
it. This changes the dynamical state of the ISM in two important ways. First,
the SMBH driven wind does not transfer angular momentum to the gas, but
increases its gravitational energy. This results in a \emph{decrease} of the
typical pericentric radius of the gas. Second, gas with differing angular momenta is pushed together, leading to (partial) cancellation.

When a black--hole accretion episode ends, the outward thrust supporting the
gas against gravity drops, and it must fall back from the radius $R_{\mathrm{shell}}$ of the swept--up region. Clearly, this infall is unlikely to be spherically symmetric. Instead, individual clumps or high--density regions fall on ballistic orbits. Because of the cancellation of angular momentum and the increase of gravitational energy during the outflow phase, these orbits are highly eccentric with pericenters much closer to
the hole than the radii from which the gas was originally swept up during the
wind feedback phase. On such eccentric orbits, any gas cloud is likely to be
tidally stretched, forming a stream, in particular near pericenter.

\begin{figure}
  \begin{center}
	\resizebox{85mm}{!}{\includegraphics{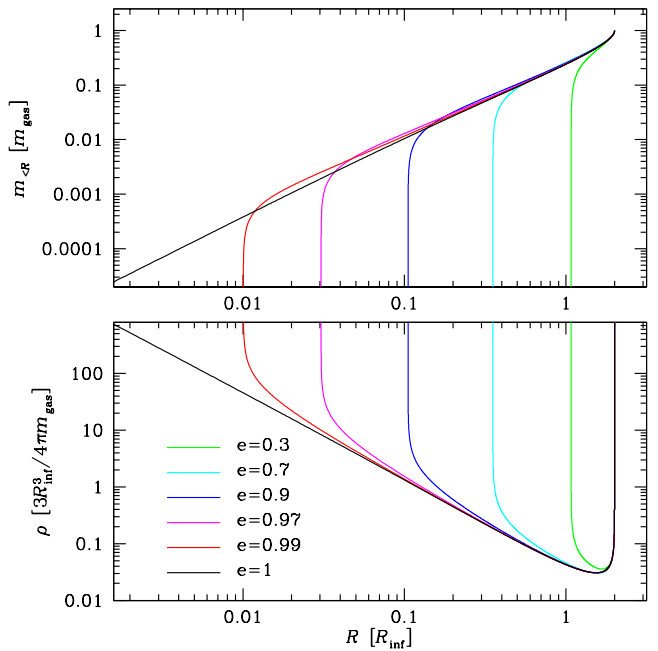}}
	\caption{Enclosed mass (\emph{top}) and mean density (\emph{bottom}) of
      a population of clouds/streams orbiting the hole with the same apocentric radius
      $\apo{R}=2R_{\mathrm{inf}}$ (corresponding to $M\approx M_\sigma/2$, for
      other choices the picture is very similar) but different eccentricities
      $e=(\apo{R}-\peri{R})/(\apo{R}+\peri{R})$ as indicated. The bulge was modeled as 	  an isothermal sphere.}
    \label{fig:rho}
  \end{center}
\end{figure}

We now estimate the resulting density of clouds/streams on such orbits. Consider a population of clouds/streams orbiting with the same peri-- and apocentric radii $R_{\scriptscriptstyle\pm}$, and hence with the same orbital energy and
specific angular momentum
\begin{equation}
  E=\frac{\apo{R}^2\apo{\Phi}^{}-\peri{R}^2\peri{\Phi}^{}}{\apo{R}^2-\peri{R}^2},
  \qquad
  L^2=\frac{2\apo{R}^2\peri{R}^2(\apo{\Phi}^{}-\peri{\Phi}^{})}{\apo{R}^2-\peri{R}^2}.
\end{equation}
Here, $\Phi_{\scriptscriptstyle \pm}\equiv\Phi(R_{\scriptscriptstyle \pm})$, where $\Phi(R)=-GMR^{-1}+\Phi_{\mathrm{bulge}}(R)$ is the total gravitational
potential. Neglecting collisions and internal shocks, the phase--space density of clouds/streams is conserved and simply the product of delta functions in $E$ and $L^2$. Integrating it over all velocities yields the spatial density
\begin{equation}
  \label{eq:rho}
  \rho(R)=\frac{m\,C}{R\sqrt{2R^2(E-\Phi( R ))-L^2}}
\end{equation}
with
\begin{equation}
  C^{-1}\equiv4\pi\int_{\peri{R}}^{\apo{R}}\frac{R\;\d R}{\sqrt{2R^2(E-\Phi(R ))-L^2}},
\end{equation}
where $m$ is the total gas mass. We identify the apocenter with the radius of the initially swept--up shell, $\apo{R}=R_{\mathrm{shell}}$, and numerically evaluate $C$ and the mass $m_{<R}=4\pi\int_{\peri{R}}^R \!\rho\,R^2\d R$ enclosed at any time within radius $R$. The resulting density and enclosed mass are plotted in Fig.~\ref{fig:rho} for various pericenters but with the apocenter fixed at $\apo{R}=2R_{\mathrm{inf}}$ with
\begin{equation} \label{eq:Rinf}
  R_{\mathrm{inf}}\equiv GM/\sigma^2
\end{equation}
the radius of the hole's sphere of influence. Because eccentric orbits have a long
residence time near apocenter, the density is maximal there and most of the gas is now further from the hole than before, in a kind of thick shell near $R_{\mathrm{shell}}$.
However, the infalling gas creates a second density maximum near pericenter, where the clouds/streams tend to collide with probability $\propto\rho^2$ and with significant relative velocity. Near apocentre, on the other hand, collisions are not only less likely (because the orbiting clouds simply return near to their initial position, avoiding each other) but also have modest relative velocities and thus do not lead to cancelation of angular momentum.

These high-impact-velocity collisions near pericentre (which are neglected in Fig.~\ref{fig:rho}) lead to accretion--disk formation because the gas loses energy much faster than angular momentum, a process familiar from accretion in close binary systems. The colliding gas must shock and lose much of its orbital energy to cooling. In addition, the collisions may cancel some, potentially most, of the angular momentum, creating a cascade of ever smaller but less eccentric orbits. Ultimately, gas on the innermost orbits circularises and forms a disk. If more gas penetrates to this radius, the disk is destroyed but quickly replaced by an even smaller one. Moreover, any misalignment of the disk angular momentum with the black hole spin results in disk tearing, when angular--momentum cancellation leads to a further reduction of the inner disk radius by a factor $10-100$ \citep{NixonKingPriceFrank2012}.

This whole process is rather complex and chaotic, but certainly has the potential to transfer some of the gas from $R_{\mathrm{gas}}\sim10-100$\,pc into an accretion disk at $R_{\mathrm{disk}}\sim0.001-0.01$\,pc, where standard viscosity--driven accretion physics takes over the mass transport, and feeds the SMBH on a timescale of $\sim10^6$\,yr. 

%%%%%%%%%%%%%%%%%%%%%%%%%%%%%%%%%%%%%%%%%%%%%%%%%%%%%%%%%%%%%%%%%%%%%%%%%%%%%%%%
\section{The Feeding Rate}
The fundamental feature of our picture is that once central accretion (and
hence feedback) slows, gas is no longer supported against gravity. This
suggests that during the chaotic infall phase, gas feeds a small--scale accretion disk
around the SMBH at some fraction of the dynamical infall rate
\begin{equation}
  \label{eq:dM:dyn}
  \dot{M}_{\mathrm{feed}}\lesssim\dot{M}_{\mathrm{dyn}}\simeq
      \frac{f_{\mathrm{g}}\,\sigma^3}{G}.
\end{equation}
For an SMBH with mass $M$
close to $M_\sigma$, this exceeds the Eddington accretion rate $\dot{M}_{\mathrm{Edd}}$ by factors $\sim10-100$ at most \citep{King2007:IAU}.

This feeding rate should characterise the rapid growth phases for the SMBH. For gas fractions $\gtrsim 0.1$ it implies disk feeding at rates a few times $\dot{M}_{\mathrm{Edd}}$. This is likely to result in the following scenario (cf.\ King \& Pringle 2006; 2007). The outer parts of the disk may become self--gravitating and form stars, while the remaining gas flows inwards under the disk viscosity at slightly super--Eddington rates. This leads to SMBH accretion at about $\dot{M}_{\mathrm{Edd}}$,
and similar mass outflow rates, with momentum scalars $\dot M_{\rm
  out}v \simeq L_{\mathrm{Edd}}/c$ \citep{KingPounds2003}. This fits
self--consistently with the feedback needed to give the observed $M-\sigma$
scaling relation \citep{King2003,King2005}.

Once central accretion stops, the SMBH should be quiescent for the sum of
the infall timescale $\apo{R}/\sigma$ and the viscous timescale
(\ref{eq:t:visc}). In general infall is more rapid, so the controlling
timescale is probably viscous and depends critically on
the radius $R_{\mathrm{disk}}$ at which the chaotic infall process places the disk.

We note that in our picture, both the precise value of the mass feeding rate
and its duty cycle are determined by essentially stochastic processes. This
makes it difficult to go beyond the simple estimates given here either
analytically or numerically. We return to this problem in the last section.

%%%%%%%%%%%%%%%%%%%%%%%%%%%%%%%%%%%%%%%%%%%%%%%%%%%%%%%%%%%%%%%%%%%%%%%%%%%%%%%%
\section{black hole obscuration}
We expect this same mechanism to produce the putative accretion `torus' at
radii larger than $R_{\mathrm{disk}}$. This structure is postulated
\citep{AntonucciMiller1985,Antonucci1993} to cover a large solid angle,
obscuring the hole along many lines of sight, and so accounting for the
populations of unobscured (Type I) and obscured (Type II) active galactic
nuclei. The main problem in understanding the torus in physical terms is that
it must consist of cool material, which by its nature cannot form a vertically
extended disk or torus. However, a large solid angle is natural if much of this
obscuring gas is not yet settled into a disk, but still falling in on a range
of orbits of very different inclinations. The column density $\Sigma=\!\int\!\rho\,\d R$ of a population of gas clouds/streams with total mass $m$ and common apo-- and pericentric radii is
\begin{equation}
	\Sigma\sim\frac{m}{2\pi(\peri{R}+\apo{R})\sqrt{\peri{R}\apo{R}}}.
\end{equation}
(using equation~(\ref{eq:rho}) with $\Phi_{\mathrm{bulge}}=0$). This diverges for small pericentric radii $\peri{R}$, so the black hole must be obscured either completely or, more probably, for many lines of sight and/or extended periods of time. In fact, the obscuring matter may not be in form of a torus at all but merely a collection of clouds/streams orbiting the hole on eccentric orbits.

Whatever the geometry of the obscuring matter, our model
renders the standard geometrical explanation for AGN unification (Antonucci 1993)
time--dependent, since the orientation of that matter changes
randomly over time and because we expect cyclicly recurring inflow phases.
This is in line with observational evidence of occasional
changes between Seyfert types \citep[e.g.][]{AlloinEtAl1985,ShappeeEtAl2013}.

%%%%%%%%%%%%%%%%%%%%%%%%%%%%%%%%%%%%%%%%%%%%%%%%%%%%%%%%%%%%%%%%%%%%%%%%%%%%%%%%
\section{The central bubble}
Our discussion so far has not specified the physical scale $R_{\mathrm{shell}}$ where the momentum--driven outflows are typically halted. Our feeding mechanism works
independently of this scale, but it may set the duty cycle and orientation of
the individual accretion disk episodes. We note that \cite{KingPounds2013}
have recently suggested that radiation pressure from the central active
nucleus tends to create a shell of gas at a characteristic radius
$R_{\mathrm{tr}}\sim 50\,(\sigma/200\mathrm{km\,s^{-1}})^2$\,pc, at which the
gas becomes transparent to the radiation from the accretion disk.

This is larger than the radius~(\ref{eq:Rinf}) of the sphere of influence by a factor
$M_\sigma/M$ and the shell's mass is comparable with the final mass $M_\sigma$ of the hole. In this picture, momentum--driven outflows must be halted here, as their inertia is of course far smaller. This means that $R_{\mathrm{shell}}\simeq R_{\mathrm{tr}}$. This idea agrees with observations of warm absorbers, which can be interpreted as arrested momentum--driven outflows.

%%%%%%%%%%%%%%%%%%%%%%%%%%%%%%%%%%%%%%%%%%%%%%%%%%%%%%%%%%%%%%%%%%%%%%%%%%%%%%%%
\section{Discussion}
We have suggested that black hole feeding is ultimately caused by feedback. By
elongating the gas orbits and promoting collisions, this causes cancellation
of opposed specific gas angular momenta, allowing accretion disks to form at
small distances from the black hole, where they can feed the hole on
time scales close to \cite{Salpeter1964}. This is different from
a situation where the gas is initially pressure supported, when cooling 
and collisions of the resulting condensations can lead to turbulent infall
\citep{GaspariRuszkowskiOh2013}. Our picture explains a number of other aspects.

As we have shown above, a near-toroidal topology for obscuring gas is a natural
result. It is also clear that the orientation of the accretion structure (disk
+ `torus') cannot be constant over time, but must be essentially random. This
is just the situation envisaged in the picture of chaotic accretion suggested
by \cite{KingPringle2006, KingPringle2007}, which results in relatively low
black hole spins. This implies rapid mass growth and low gravitational--wave
recoil velocities for merging black holes. The impact of the black hole wind
on the gas which ultimately falls in may cause some of it to form stars, and
this can also happen in the collisions during gas infall. Of course, any gas converted to and/or heated by stars is prevented from participating in the black hole feeding. However, at each feeding cycle only a small fraction of the gas within $R_{\mathrm{shell}}$ is required to reach $R_{\mathrm{disk}}$, and only gas locked in stellar remnants and dwarfs is ultimately prevented from accreting.

Because angular momentum has been largely cancelled, such newly formed
stars fall in on near-parabolic orbits. This has several consequences.
First, stars coming too close to the hole create visible tidal disruption
events \citep{Rees1988}; second, tidal dissociation of close binaries produces hypervelocity stars \citep{Hills1988}; finally, massive stars which escape these fates inject metal--enriched gas into their surroundings. In any plausible picture most of
this gas remains near to the hole, and could undergo repeated star
formation. This may be the origin of the high chemical enrichment observed in
AGN spectra \citep{Shields1976,BaldwinNetzer1978,
  HamanFerland1992,FerlandEtAl1996,DietrichEtAl1999,DietrichEtAl2003:AaA,
  DietrichEtAl2003:ApJ,AravEtAl2007}.

We note that the idea of feedback--stimulated feeding opens the possibility of
runaway growth: the black hole forages for its own food, and grows still
faster. Given an abundant food supply (i.e. $f_{\mathrm{g}} \gtrsim 0.1$) this growth is
stopped only as the hole reaches the limiting $M - \sigma$ mass and drives all
the food away. A runaway SMBH like this would of course have a tendency to
grow at the Eddington rate for most of its (short) feeding frenzy. This may
explain very massive SMBH observed at high redshifts
\citep[e.g.][]{BarthEtAl2003, WillottMcLureJarvis2003, FanEtAl2003,
  MortlockEtAl2011}. Here the close proximity of all galaxies means that many
are likely to be gas--rich (i.e. $f_{\mathrm{g}} \gtrsim 0.1$) because of mergers, so
runaways are favored.

One interesting aspect of the proposed mechanism is the mutual dependence of
feeding and feedback on each other. Clearly, this whole process must be started by some initial accretion which was \emph{not} triggered by feedback, but by sufficient gas coming within $\lesssim0.001\,$pc of the infant hole. Such an event could be triggered by a galactic merger, but must be relatively rare. This implies that early SMBH formation may be somewhat random, but more likely in frequently perturbed/merging galaxies.

Conversely, if the SMBH's neighbourhood at $R\lesssim R_{\mathrm{gas}}$ acquires some net rotation, for example, during a merger, then the distribution of
angular momenta is unlikely to allow for angular-momentum cancellation. In such a situation, the SMBH suffers from starvation. Despite sitting tantalisingly close to its food, it cannot reach it nor bring it down easily. However, if the rotating gas can cool, it will form a disk (and possibly stars), clearing most of the space and opening the possibility for re-starting the feeding cycle.

Also, our proposed feeding mechanism will not work efficiently if the feedback  is dominated by a collimated jet rather than wide--angle outflows. This is obvious if the impact shocks are efficiently cooled (momentum--driven flow) as the jet simply carves a narrow hole in the gas it impacts. If the shocks do not cool (energy--driven), their effect is wider but still unlikely to cause feeding in the way described here.

Finally, we note that the episodic in-- and outflows of a fraction $f_{\mathrm{g}}\sim0.1$ of matter at velocities well above $\sigma$ entail abrupt changes in the gravitational potential in the inner $R_{\mathrm{shell}}\sim10-100\,$pc. Therefore, the growth of the SMBH is not an adiabatic process for the dynamics of collisionless matter on these scales. Instead, the abrupt variations in the potential redistribute the orbital actions. This renders the central dark--matter density smaller than current estimates \citep[by e.g.][]{Young1980, QuinlanHernquistSigurdsson1995} based on the adiabatic assumption, though possibly still larger than in absence of a SMBH. This  implies a significant reduction in the expected dark--matter annihilation signal
from SMBH hosting galaxy centers.

%%%%%%%%%%%%%%%%%%%%%%%%%%%%%%%%%%%%%%%%%%%%%%%%%%%%%%%%%%%%%%%%%%%%%%%%%%%%%%%%
\section{A subgrid recipe}
We have suggested that feeding of supermassive black holes may in many cases
be stimulated by feedback. A practical question is how one might implement
this process in simulations of galaxy formation which cannot resolve the hole's
sphere of influence, let alone the dynamics and cooling of infall and outflow, and instead must use a subgrid recipe. Clearly, any Bondi-like subgrid recipe adapted to account for the angular momentum of the gas at $\gtrsim R_{\mathrm{gas}}$ cannot adequately describe these dynamics. Instead a completely different approach is required.

We have seen that feedback-induced feeding generally occurs at a fraction of the dynamical infall rate (\ref{eq:dM:dyn}) when it operates. This is generally slightly super--Eddington (for $f_{\mathrm{g}} \gtrsim 0.1)$. This in turn makes the SMBH grow at about the Eddington rate, and rejects the remainder of the mass in a wind, which is what causes the feedback. If $M<M_\sigma$ we know in reality
this will result in momentum--driven feedback, which keeps the accretion
going, and does not blow the gas away. Once $M\ge M_\sigma$, the feedback
changes character to energy--driven and terminates SMBH mass growth.

Given the discussion above, a suitable subgrid recipe is as follows. Grow $M$
from surrounding gas at the rate $\dot{M}=\min\{\epsilon \dot M_{\mathrm {dyn}}, \dot M_{\mathrm{Edd}}\}$ (see equation~\ref{eq:dM:dyn}) with $\epsilon\sim0.1$. If $M< M_\sigma$, \emph{neglect} feedback. If $M\ge M_\sigma$, deposit \emph{energy} into the
surrounding gas at the rate $(\eta/2) c^2\dot{M}\simeq0.05c^2\dot{M}$ \citep{ZubovasKing2012:ApJ}.

%%%%%%%%%%%%%%%%%%%%%%%%%%%%%%%%%%%%%%%%%%%%%%%%%%%%%%%%%%%%%%%%%%%%%%%%%%%%%%%%
\section*{Acknowledgments}
We thank Ken Pounds, Chris Nixon, and Peter Hague for helpful conversations.
Theoretical astrophysics in Leicester is supported by an STFC Consolidated Grant.

%%%%%%%%%%%%%%%%%%%%%%%%%%%%%%%%%%%%%%%%%%%%%%%%%%%%%%%%%%%%%%%%%%%%%%%%%%%%%%%
%\bibliographystyle{mnras}
%\bibliography{refs} 
%\end{document}
%%%%%%%%%%%%%%%%%%%%%%%%%%%%%%%%%%%%%%%%%%%%%%%%%%%%%%%%%%%%%%%%%%%%%%%%%%%%%%%%

%%%%%%%%%%%%%%%%%%%%%%%%%%%%%%%%%%%%%%%%%%%%%%%%%%%%%%%%%%%%%%%%%%%%%%%%%%%%%%%%
\end{document}